# Studies of Multiferroic System LiCu$_2$O$_2$ I
## Sample Characterization and Relationship between Magnetic Properties and Multiferroic Nature

Yukio Yasui, Kenji Sato, Yoshiaki Kobayashi, and Masatoshi Sato[*]

*Department of Physics, Division of Material Science, Nagoya University, Furo-cho, Chikusa-ku, Nagoya 464-8602*

Single-crystal samples of LiCu$_2$O$_2$ with spin 1/2 Cu$^{2+}$ chains of edge-sharing CuO$_4$ square planes (ribbon chains), have been characterized by X-ray diffraction, thermogravimetric analysis, and magnetic measurements. Neither the atomic deficiency nor the mixing of Cu and Li atoms has been found, indicating that lattice defects conceived as a possible origin of the reported multiferroic behavior can be excluded. Anomalies found in the data of specific heat and neutron magnetic Bragg reflections show clear evidence that the system exhibits successive magnetic transitions at $T_{N1}$~24.5 K and $T_{N2}$~22.8 K. Based on the magnetic structures in the intermediate ($T_{N2}<T<T_{N1}$) and low temperature ($T<T_{N2}$) phases, determined by the combined studies of neutron scattering and $^7$Li-NMR measurements, we can consistently understand the fact that the multiferroic properties are observed only below $T_{N2}$ by considering existing theories.

*corresponding author: e43247a@nucc.cc.nagoya-u.ac.jp

## 1. Introduction

Quasi one-dimensional Cu$^{2+}$ $S$=1/2 spins in the so-called ribbon chains of edge-sharing CuO$_4$ square planes attract much attention. For the spins, the exchange interaction between the next-nearest-neighbor spins $J_2$ (NNN) is antiferromagnetic ($J_2$>0) and nearest-neighbor (NN) exchange interaction $J_1$ often becomes ferromagnetic (FM; $J_1$<0) or has a rather small value even when it is antiferromagnetic (AFM; $J_1$>0). This causes the competition of these interactions and induces nontrivial magnetic structures. Theoretically, if $|J_2/J_1|$ is larger than a critical value, the helical magnetic structure is expected.[1,2] Actually, LiVCuO$_4$, LiCu$_2$O$_2$, Li$_2$ZrCuO$_4$ *etc*. were reported, for examples, to have such helical magnetic structures.[3-10]

A renewed interest has been focused on the helimagnet systems[11-14] from the view point of multiferroics, which have the magnetic and ferroelectric simultaneous transitions, and theoretically, it has been pointed out that the helical magnetic structure induces the multiferroic nature.[13-16] Recently, the authors' group have found that LiVCuO$_4$ is a multiferroic, where the relation $\boldsymbol{P} \propto \boldsymbol{Q} \times \boldsymbol{e}_3$ is found,[4-6] consistently with the theories,[15-18] where $\boldsymbol{P}$, $\boldsymbol{Q}$ and $\boldsymbol{e}_3$ are the ferroelectric polarization, the modulation vector and the helical axis of the ordered spins, respectively. To understand the multiferroic nature, studies of the $S$=1/2 multiferroic systems are very useful, because there are not any complications due to the multi-orbital effects. Moreover, these $S$=1/2 multiferroic systems are expected to exhibit the interesting phenomena induced by quantum effects.

LiCu$_2$O$_2$ is one of the examples of systems with the CuO$_2$ ribbon chains and its structure is shown schematically in Fig. 1. It is orthorhombic (space group *Pnma*) and there are four ribbon chains along the *b* direction in a unit cell.[19] The chains are separated by the nonmagnetic Li$^+$ and Cu$^+$ ions. For this system, multiferroic nature was reported by Park *et al*.[20] The data indicate that the occurrence of the ferroelectricity is closely related to the magnetic transition to a nontrivial magnetic structure. However, there are many unsolved problems for the details of the multiferroic behavior: For example, although several groups studied the magnetic structure of LiCu$_2$O$_2$ by means of neutron scattering,[6,8] $^7$Li-NMR,[7] resonant soft x-ray magnetic scattering,[21,22] *etc.*, it has not been completely determined yet. Moreover, it seems to be still unclear if the relation $\boldsymbol{P} \propto \boldsymbol{Q} \times \boldsymbol{e}_3$ holds or not. To clarify these problems, we have carried out systematic studies of the systems, including the preparations of the samples, their characterizations, various kinds of measurements of macroscopic physical quantities, and neutron scattering and NMR studies as the microscopic probes.

In the present paper (paper I), results of the sample characterizations and various macroscopic measurements are mainly presented in relation to the magnetic structure determined by the combined works of the neutron and NMR measurements. (We report the detailed determination of its magnetic structure by the companion paper, paper II).[23] Powder X-ray diffraction studies, thermogravimetric analysis (TG) and measurements of magnetic properties have been carried out on single-crystal samples of LiCu$_2$O$_2$ to characterize the samples. Much effort has been made to clarify if the lattice imperfections and chemical disorder exist or not in the samples.

Two anomalies have been observed in the $T$-dependence of the specific heat $C_s$ and neutron magnetic scattering intensity, which ensures the existence of two magnetic transitions at $T_{N1}$~24.5 K and $T_{N2}$~22.8 K reported in ref. 8. At these temperatures, we have found that the dielectric susceptibility ε measured with the electric field $\boldsymbol{E}$ along *c* has small but clear anomalies. Based on these results and those of the magnetic structure determination described in paper II in detail, the relationship between the magnetic structure and dielectric properties is discussed. Possible effects of the quantum nature of the Cu$^{2+}$ spins are also discussed.

## 2. Experiments

Single-crystal samples of LiCu$_2$O$_2$ were prepared basically by the method reported in ref. 24: Mixtures of Li$_2$CO$_3$ and CuO powders with the molar ratio 1:4.2 were heated to 1150 ºC at a rate of 200 ºC/h, kept at the temperature for 10 h, cooled down to 900 ºC at a rate of 10 ºC/h and then quenched to room temperature. The Zn doped single-crystals (LiCu$_{2-x}$Zn$_x$O$_2$) have been grown by a similar method, where the initial mixtures of Li$_2$CO$_3$, CuO and ZnO with a molar ratio 1.2: 4-2$x$: 2$x$ were used. The obtained crystals were confirmed not to have appreciable amounts of impurity phases by X-ray measurements on the pulverized samples. The crystal axes were determined by observing the X-ray diffraction peaks. The typical size of the crystals was 15×15×1mm$^3$. By a polarized optical microscope, the existence of the twin structure was observed, which is due to the fact that the lattice parameter $a$ is very close to 2$b$.[19]

The magnetic susceptibilities χ and the magnetization $M$ were measured using a SQUID magnetometer in the temperature $T$ range from 2 to 300 K in the magnetic field up to 5.5 T. The specific heat $C_s$ was measured by the thermal relaxation method using a Quantum Design PPMS. To study the dielectric susceptibility ε, the capacitance $C$ of the rectangular sample plates with dimensions of 3.0×2.0×0.7 mm$^3$ was measured with the electric field $E$//$c$, where the electrodes were attached with silver paint to both sides of the plate surface. An ac capacitance bridge (Andeen Hagerling 2500A) with a frequency of 1 kHz was used. If the stray capacitance is negligible, dielectric susceptibility ε is directly proportional to the observed capacitance $C$.

Neutron measurements on a single crystal were carried out using the triple spectrometer TAS-1 installed at JRR-3 of JAEA in Tokai. To avoid the large neutron absorption of Li, we used $^7$Li isotope in the growth of crystals for neutron scattering. The obtained single-crystal samples were ground to powder, and the powder X-ray diffraction pattern was taken at room temperature with the Cu $Ka$ radiation to carry out the Rietveld analyses by using the computer program Rietan 2000.[25] The thermogravimetric analyses (TG) with use of RIGAKU-TG have been carried out in the flowing O$_2$ and flowing He with 5 % H$_2$ (100 cm$^3$/min.).

**3. Results and Discussion**

The $T$-dependence of the specific heat divided by $T$, $C_s/T$ of LiCu$_2$O$_2$ is shown in Fig. 2(a), where the two anomalies are clearly observed at temperatures $T_{N1}$~24.5 K and $T_{N2}$~22.8 K, confirming the results of ref. 8 obtained by the magnetic and dielectric susceptibility measurements. Figure 2(b) shows the neutron magnetic scattering intensities of 1/2 1-δ 0 reflection of LiCu$_2$O$_2$ (δ~0.172). From the figure, we find that the magnetic ordering grows with decreasing $T$ below $T_{N1}$, and that the intensity-$T$ curves exhibits an anomaly at $T_{N2}$, indicating the change of the magnetic structure at this temperature. The behavior of the $^7$Li-NMR spectra (not shown here)[23] also supports that the magnetic structure in the intermediate phase is different from that below $T_{N2}$. Figure 2(c) shows the $T$-dependence of the capacitance $C$ measured with the electric field $E$//$c$ of LiCu$_2$O$_2$, where small but clear peaks of the capacitance can be found at $T_{N1}$ and $T_{N2}$. The relationship between the magnetic structures in the two phases and the dielectric properties will be discussed later.

The $T$-dependences of the magnetic susceptibilities χ of LiCu$_2$O$_2$ measured under the condition of the zero field cooling (ZFC) are shown in the inset of Fig. 3(a) for the applied field $H$//$a$ and $H$//$c$. The broad peaks of χ at ~35 K can be attributed to the growth of the short-range spin correlation with decreasing $T$. The temperature derivative of χ, $d$χ$/dT$ (not shown here) indicates that the susceptibility χ measured with $H$ parallel to $c$ and $a$ decrease significantly with decreasing $T$ at $T_{N1}$=24.5K and at $T_{N2}$=22.8 K, respectively. With further decreasing $T$, χ becomes almost constant in the region of $T$ < 10 K, indicating that the Curie component induced by lattice imperfections and/or chemical disorders do not exist for the present single-crystal samples. The $T$-dependences of χ of the samples of LiCu$_{2-x}$Zn$_x$O$_2$ with $x$=0.05, 0.10 and 0.15 measured under the condition of ZFC for the magnetic field of 0.1 T ($H$//$ab$-plane) are also shown in Fig. 3(a). The Curie component in the $T$-region of $T$<15 K increases with increasing Zn concentration. This result indicate that the free spins are induced by the Zn substitution for Cu. (However, the number of the free spins estimated from the Curie constant is much smaller than that of the doped Zn concentration, which can be understood by the fact that the doped Zn atoms do not always induce free spins and do not cut the spin correlation. It is because the next nearest interaction is large.) These results indicate that the lattice imperfection for the present single-crystal samples of LiCu$_2$O$_2$ is not important for the arguments described below.

Figure 3(b) shows the magnetization curves $M$ of LiCu$_2$O$_2$ measured in two different field directions at 5K. The values of $M$ increase linearly with increasing magnetic fields for both directions of $H$//$a$ and $H$//$c$, indicating the nonexistence of a spin flop transition in the magnetic field up to 5.5 T. This should be contrasted with the facts that the spin flop transition is observed at $H$= ~2 T and ~4 T for LiVCuO$_4$[5,10] and Li$_2$ZrCuO$_4$[9], respectively.

The X-ray powder diffraction pattern taken at room temperature for LiCu$_2$O$_2$ is shown in Fig. 4. We have carried out Rietveld refinement using the space group $Pnma$ with keeping the stoichiometry of LiCu$_2$O$_2$, where the Li$^+$ ionic coordinates are fixed to the reported values.[19] The fitting is found to be almost satisfactory ($R_{wp}$=4.45, $S$=1.84), and the obtained structural parameters are consistent with the results reported in Ref. 19. Then, we have tried to carry out additional analyses by using the chemical formula Li$_{1-x}$Cu$_{2+x}$O$_2$, -1<$x$<1, and found the $x$ value of 0.01±0.02 ($R_{wp}$=4.45, $S$=1.84). The analyses adopting the fixed Li$_{1.16}$Cu$_{1.84}$O$_2$ formula reported in ref. 6 have also been carried out, and the values $R_{wp}$=6.63 and S=2.74 are obtained. These results also exclude effects of the lattice imperfections of Cu atoms for the single-crystal samples of LiCu$_2$O$_2$ within error bars.

In order to estimate the Li content, thermogravimetric analyses (TG) have also been carried out. The rates of the sample-mass change, Δ$m/m_0$ taken with increasing $T$ in the flowing O$_2$ (thick line) and flowing He with 5 % H$_2$ (thin line) are shown in Fig. 5, where the $m_0$ and Δ$m$ are the initial sample mass at room temperature and Δ$m$ = $m(T)$–$m_0$. For the samples in the plateau region of the Δ$m/m_0$-$T$ curves indicated by the arrows, the thermogravimetric analyses have been carried out, and found followings. In flowing O$_2$ gas, Li$_{1-y}$CuO$_2$ decomposes into Li$_2$CuO$_2$ and CuO in the region of 700 ºC<$T$<800 ºC, as described by the relation,

Li$_{1-y}$Cu$_2$O$_2$+(1-$y$)/4×O$_2$=(1-$y$)/2×Li$_2$CuO$_2$+(3+$y$)/2×CuO.   (1)

From the weight gain after the sample was completely converted into $Li_2CuO_2$ and CuO, we obtained the $y$ value of 0.01±0.05. (The value of $\Delta m/m_0$=0.0482 for $y$=0 should be compared with the observed value in the plateau region.) In flowing He with 5 % $H_2$, $Li_{1-y}CuO_2$ decomposes into $Li_2O$, CuO and Cu in the plateau region indicated by the arrow (500 ºC<$T$<600 ºC). (At $T$ higher than 600 ºC, the de-oxidization reaction of the samples is still progressing, and at 1100 ºC, the system finally decomposes to $Li_2O$ and Cu.) The chemical formula at the plateau is considered to be described as

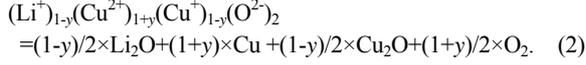

$(Li^+)_{1-y}(Cu^{2+})_{1+y}(Cu^+)_{1-y}(O^{2-})_2$
=(1-$y$)/2×$Li_2O$+(1+$y$)×Cu +(1-$y$)/2×$Cu_2O$+(1+$y$)/2×$O_2$.   (2)

We have calculated the $y$ value, using the Eq. 2, and then obtained the $y$ value of –0.04±0.05. The other reaction formula might be considered. However, the $y$-values estimated by the other formulas are too large and are unrealistic. If $y$=0, we expect $\Delta m/m_0$= –0.0964, in the plateau region indicated by the arrow, which should be compared with the observed value in the plateau region.) From these results, the Li deficiency does not exist in the present single-crystal samples of $LiCu_2O_2$ within the experimental error bars.

We have determined the magnetic structures of $LiCu_2O_2$ in the intermediate phase ($T_{N2}$<$T$<$T_{N1}$) and low temperature one ($T$<$T_{N2}$), by the combining results of neutron scattering and $^7$Li-NMR. The detailed descriptions of the magnetic-structure determination can be found in the paper II.[23] Here, we just show the magnetic structures which can reproduce both the observed magnetic scattering intensities and $^7$Li-NMR spectra of $LiCu_2O_2$ at $T$=23.3K (intermediate phase) in Fig. 6(a) and 6(c), and at 5 K (low temperature phase) in Fig. 6 (b) and 6 (d), schematically. At $T$=23.3K, spins have collinear and sinusoidal structure with the spin directions along the $c$-axis. The components of the ordered moments can be described by the equations, $m^x_i(y)$=0, $m^y_i(y)$=0 and $m^z_i(y)$=$\mu_c$cos($Q·y$+$\phi_i$), where the pitch angle $Q·b$=$\Delta\phi$~62.03 º, $\mu_c$=0.3±0.1 $\mu_B$ and $\phi_i$ describes the relative phases in the $i$-th ($i$ =1~4) $CuO_2$ chains within a unit cell, which correspond to the labels ①~④ in Fig. 6(a), respectively, and one possible set of these values are $\phi_1$=0 º, $\phi_2$=90 º, $\phi_3$=90 º and $\phi_4$=180 º, and another set $\phi_1$=0 º, $\phi_2$= -90 º, $\phi_3$= -90 º and $\phi_4$= -180 º is also possible. In Figs. 6(a) and 6(c) correspond to the former set, and the latter can be obtained by reversing the $b$-axis.

The magnetic structure of the low temperature phase is very complicated and spins have nonzero components in all of the three directions. One of the obtained magnetic structure at $T$=5 K is a kind of ellipsoidal helical structure shown in Fig. 6(b) and 6(d). The details can be described as follows: The components of the ordered moments are $m^x_i(y)$=$\mu_{ab}$sin($Q·y$+$\phi_i$)·sin$\alpha$, $m^y_i(y)$=$\mu_{ab}$sin($Q·y$+$\phi_i$)·cos$\alpha$ and $m^z_i(y)$=$\mu_c$cos($Q·y$+$\phi_i$), the modulation amplitudes $\mu_{ab}$=0.45±0.10 $\mu_B$ and $\mu_c$=0.85±0.15 $\mu_B$, and the other parameters $\alpha$= –45 º(and 135 º), $\phi_1$=0, $\phi_2$= 90 º, $\phi_3$=90 º and $\phi_4$=180 º present two magnetic patterns. The two patterns have the opposite signs of the spin-rotation axis $e_3$. Other two sets are give by the parameters, $\mu_{ab}$=0.45±0.10 $\mu_B$, $\mu_c$=0.85±0.15 $\mu_B$, $\alpha$= 45º(and –135º), $\phi_1$= 0º, $\phi_2$= –90º, $\phi_3$= –90º and $\phi_4$= –180º is also possible. These two can be obtained by reversing the $b$-axis of the former patterns. In $E$=0, the magnetic domain with the helical axis $e_3$ along [1 1 0], [1 -1 0], [-1 1 0] or [-1 -1 0] seems to exist. Spins at $Cu^{2+}$ sites shifted by (1, 0, 0) are anti-parallel to those at the original sites, producing the magnetic Bragg reflections at the points with $h$=half integer in both the intermediate and low temperature phases.

The existence of the sinusoidally modulated collinear phase above $T_{N2}$, and the observed anisotropy of the amplitudes of the modulations, which form the helical structure may be related to a possible effect of the quantum fluctuation proposed by Ref. 26.

Here, based on these results, the relationship between the magnetic structures and dielectric properties is discussed. Although the peak of the capacitance-$T$ curve is observed at $T_{N1}$, as shown in Fig. 2(c), the ferroelectricity has not been observed in the intermediate phase.[8,21] We discuss later why the peak appears at this temperature. The magnetic structure in the intermediate phase is collinear modulated structure, which is consistent with existing theories.[13-16] While, in the low temperature phase, the magnetic structure is ellipsoidal helical one and at $H$=0, and the ferroelectric polarization $P$ is expected to be along $c$ from the relation $P \propto Q \times e_3$. It is consistent with the observation.[8,21]

Here, we make a following comment. The peaks of the $\varepsilon$-$T$ curve are observed not only at $T_{N2}$ but also at $T_{N1}$, as shown in Fig. 2(c). The peak at $T_{N1}$ does not necessarily indicate, we think, the existence of the ferroelectric transition at this temperature: The electric susceptibility $\varepsilon$ at around $T_{N2}$ is expected to have the $T$ dependence shown in Fig. 6(e) by the dashed line, because the system exhibits the ferroelectric transition at $T_{N2}$. However, the increasing tendency of $\varepsilon$ is suppressed at $T_{N1}$ by the collinear ordering. Then, due to the restoration of the helical fluctuation, the anomaly can be seen at $T_{N2}$, as shown in Fig. 6(e) by the solid line. Mostovoy[16] developed the Landau expansion theory that if the system exhibits the magnetic transition to the collinear and sinusoidally modulated state, the temperature derivative of $\varepsilon$ exhibits discontinuity at the transition temperature. This discontinuity can explain the peak of $\varepsilon$ at $T_{N1}$.

Moskvin et al.[27] proposed that the ferroelectricity observed for $LiCu_2O_2$ is due to the non-stoichiometry of the system, and explained the puzzles of the dielectric properties in the magnetic field reported by Park et al.,[20] where they considered the spin-flop transition. However, we have found that the lattice imperfections, such as the atomic deficiency and the mixing of Cu and Li atoms do not exist within the experimental error bars for the present samples of $LiCu_2O_2$. Evidence for the existence of free spins induced by the non-stoichiometry has not been found. The fact that the relation $P \propto Q \times e_3$ consistent with the existing theories[15-18] holds in zero magnetic field. These results indicate that the ferroelectricity of $LiCu_2O_2$ is induced by the helical type magnetic structure. In the magnetization curves, the spin flop transition has not been observed in the magnetic field up to 5.5 T. Even by using the resent information, the dielectric properties observed by Park et al.[20] in finite magnetic fields are still difficult to explain. It remains as a future problem.

## 4. Conclusions

We have shown the results of powder X-ray diffraction studies, thermogravimetric analyses and magnetic measurements of macroscopic physical properties carried out on single-crystal samples of $LiCu_2O_2$. For the samples of $LiCu_2O_2$ prepared in the present studies, the atomic deficiency and the mixing of Cu and Li atoms have not been found within the experimental error bars. Studies by specific heat measurements and neutron magnetic scattering indicate that the system exhibits two magnetic transitions at $T_{N1} \sim 24.5$ K and $T_{N2} \sim 22.8$ K. Based on the obtained magnetic structures determined in the intermediate ($T_{N2}<T<T_{N1}$) and low temperature ($T<T_{N2}$) phases, the relationship between the magnetic structures and dielectric properties has been argued: In the intermediate phase, the system has the collinear magnetic structure with the sinusoidal modulation and the ferroelectricity is not induced, while in the low temperature region, it has the ellipsoidal helical structure and the relation $\boldsymbol{P} \propto \boldsymbol{Q} \times \boldsymbol{e}_3$ has been found to hold, consistently with the theories derived by the phenomenological[16] and microscopic[15,17,18] models.

**Acknowledgments**

This work is supported by Grants-in-Aid for Scientific Research from the Japan Society for the Promotion of Science (JSPS) and by Grants-in-Aid on Priority Area from the Ministry of Education, Culture, Sports, Science and Technology.

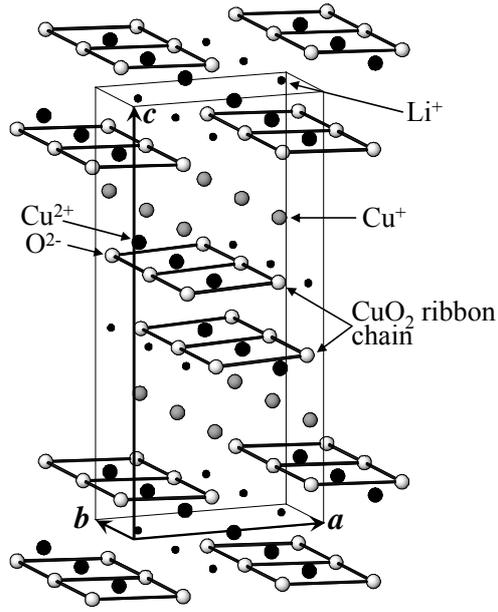

Fig. 1. The schematic structure of LiCu$_2$O$_2$. The one-dimensional chains of the edge sharing CuO$_4$ squares (CuO$_2$ ribbon chains) run along *b*, which are separated by the nonmagnetic Li$^+$ and Cu$^+$ ions.

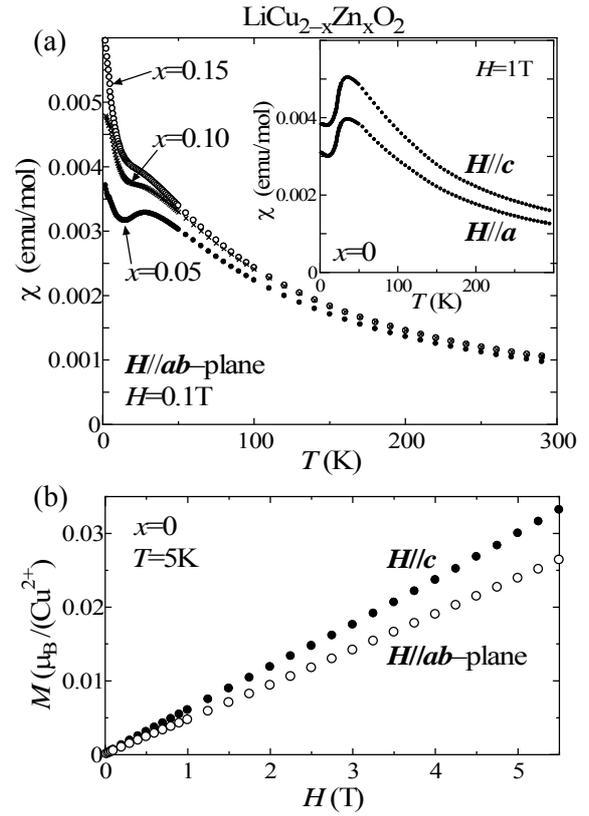

Fig. 3. (a) *T*-dependence of the magnetic susceptibilities χ of LiCu$_{2-x}$Zn$_x$O$_2$ for *x*=0.05, 0.10 and 0.15 measured under the condition of the zero field cooling for the magnetic field of 0.1 T (*H*//*ab*-plane). Inset shows the *T*-dependence of χ of a single crystal sample of LiCu$_2$O$_2$ for two different field directions. (b) Magnetization curves of LiCu$_2$O$_2$ are shown for two different field directions at 5K.

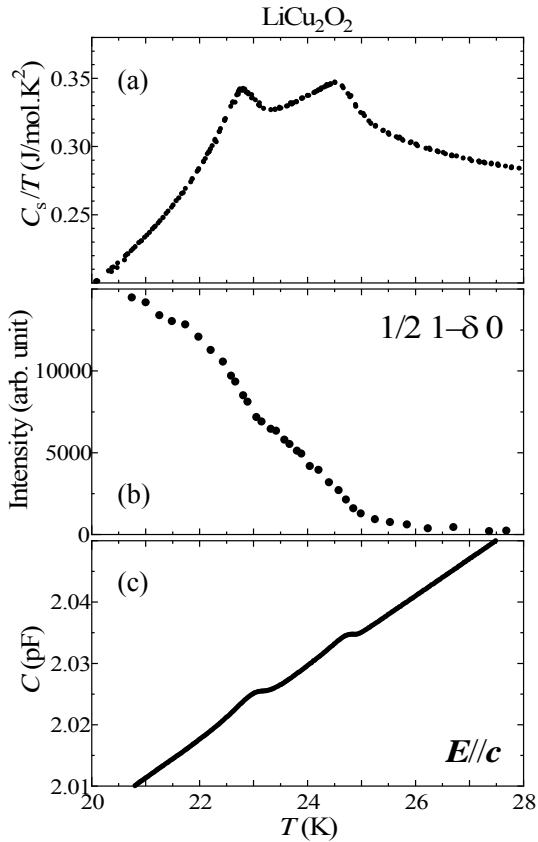

Fig. 2. (a) specific-heat divided by *T*, (b) neutron magnetic scattering intensities of 1/2 1-δ 0 (δ~0.172) reflection and (c) capacitance measured for the electric field *E*//*c* are shown against *T*.

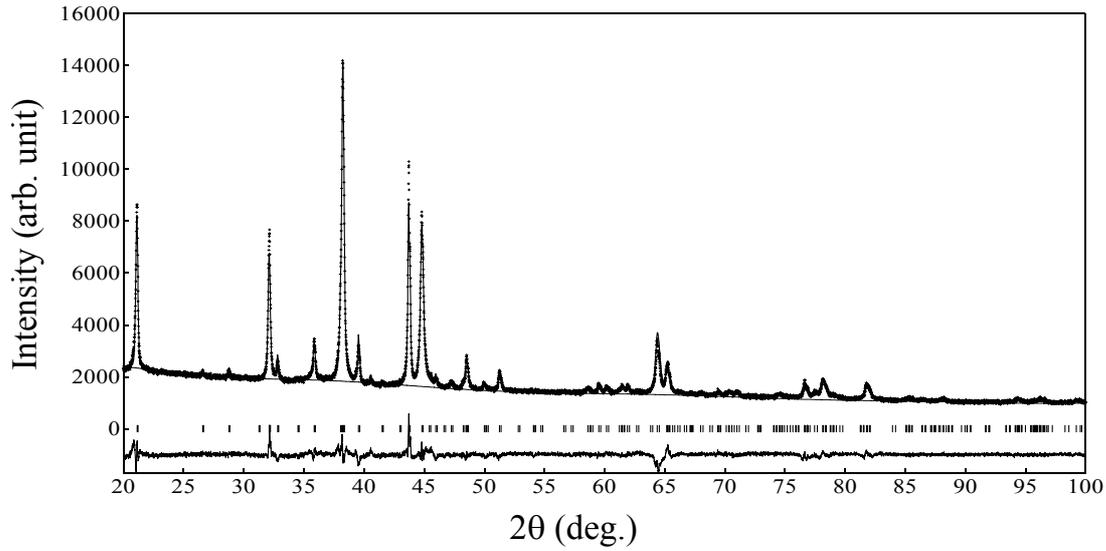

Fig. 4. Powder X-ray diffraction pattern (dots) observed at 300K is shown together with the fitted curve of the Rietveld refinement (solid line) for $LiCu_2O_2$.

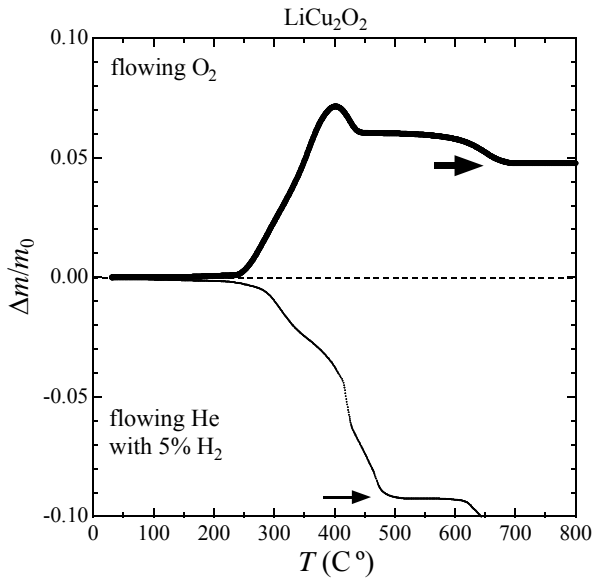

Fig. 5. The thermogravimetric (TG) curves taken with varying $T$ in the flowing $O_2$ (thick line) and flowing He with 5 % $H_2$ (thin line) are shown. In the plateau regions indicated by the arrows, the $\Delta m/m_0$-$T$ curves are analyzed in order to estimate the Li content of the single crystal samples of $LiCu_2O_2$ (see the text for details.)

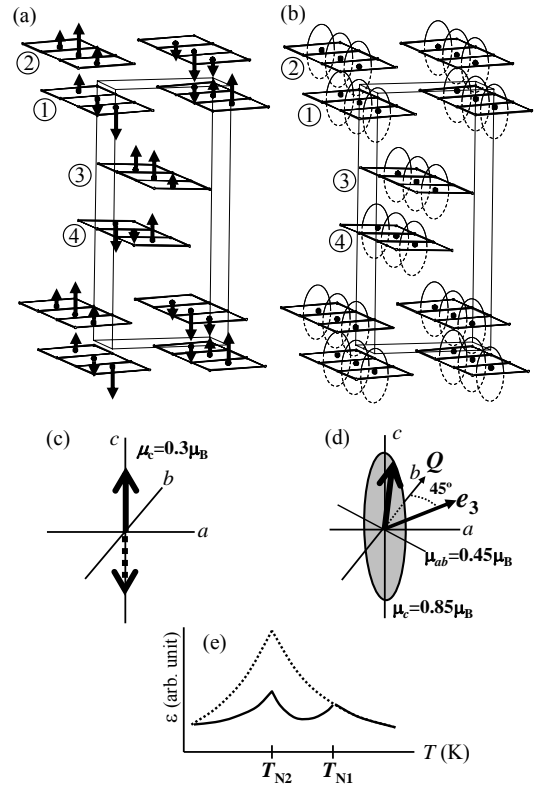

Fig. 6. (a) and (b) Magnetic ordering patterns which can reproduce the magnetic scattering intensities and $^7$Li-NMR spectra of $LiCu_2O_2$ at (a) $T$=23.3K (intermediate phase) and (b) $T$=5K (low temperature phase). In (b) only the rotating planes of the $Cu^{2+}$ spins are shown. Details of analyses of magnetic structures are described in ref. 23. The figures (c) and (d) show the detailed parameters describing the magnetic structures in (a) and (b), respectively. The thick arrows indicate the direction of the $Cu^{2+}$ moments, and $Q$ and $e_3$ are the directions of the modulation vectors and helical axis, respectively. (e) Schematic $\varepsilon$-$T$ curves at around two transition temperatures.